\documentclass[conference]{IEEEtran}

\usepackage{xspace}
\usepackage{graphicx}
\usepackage{pifont}
\usepackage{algorithmic}
\usepackage[algoruled]{algorithm2e}
\usepackage{float}
\usepackage{moreverb}
\usepackage{booktabs}
\usepackage{multirow}
\usepackage{subfigure}
\usepackage{color}
\usepackage{etoolbox}
\usepackage{arydshln}
\usepackage{caption}
\usepackage{url}

\graphicspath{{.}{./figure/}}

\newcommand{\xref}[1]{Section~\ref{#1}}

\newcommand{\fref}[1]{Fig.~\ref{#1}}
\newcommand{\tref}[1]{Table~\ref{#1}}

\newcommand{\first}{First,~}
\newcommand{\second}{Second,~}
\newcommand{\third}{Third,~}

\newcommand{\ie}{i.\,e., \@}
\newcommand{\eg}{e.\,g., \@}

\newcommand{\etal}{et~al.\xspace}
\newcommand{\perc}{\,\%\xspace}

\newcommand{\wid}{\textsf{\begin{footnotesize}World IPv6 Day\end{footnotesize}}\xspace}
\newcommand{\wil}{\textsf{\begin{footnotesize}World IPv6 Launch\end{footnotesize}}\xspace}
\newcommand{\teredo}{\textsf{\begin{footnotesize}teredo\end{footnotesize}}\xspace}

\newcommand{\ixpa}{IXP11-v6day\xspace}
\newcommand{\ixpb}{IXP12-newyear\xspace}
\newcommand{\ixpc}{IXP12-launch\xspace}
\newcommand{\ixpd}{IXP12-olympics\xspace}

\begin{document}

\title{Watching the IPv6 Takeoff from an IXP's Viewpoint}

\author{
\IEEEauthorblockN{Juhoon Kim}
\IEEEauthorblockA{TU-Berlin\\
Berlin, Germany\\
jkim@net.t-labs.tu-berlin.de}
\and
\IEEEauthorblockN{Nadi Sarrar}
\IEEEauthorblockA{TU-Berlin\\
Berlin, Germany\\
nadi@net.t-labs.tu-berlin.de}
\and
\IEEEauthorblockN{Anja Feldmann}
\IEEEauthorblockA{TU-Berlin\\
Berlin, Germany\\
anja@net.t-labs.tu-berlin.de}}

\maketitle

\begin{abstract}
The different level of interest in deploying the new Internet address
space across network operators has kept IPv6 tardy in its deployment.
However, since the last block of IPv4 addresses has been assigned, Internet
communities took the concern of the address space scarcity seriously and started
to move forward actively. After the successful IPv6 test on 8 June, 2011
(\wid~\cite{URL-V6DAY}), network operators and service/content providers were
brought together for preparing the next step of the IPv6 global deployment
(\wil on 6 June, 2012~\cite{URL-V6LAUNCH}). The main purpose of the event was
to permanently enable their IPv6 connectivity.

In this paper, based on the Internet traffic collected from a large European
Internet Exchange Point (IXP), we present the status of IPv6 traffic mainly
focusing on the periods of the two global IPv6 events. Our results show that
IPv6 traffic is responsible for a small fraction such as 0.5\perc of the total
traffic in the peak period. Nevertheless, we are positively impressed by the
facts that the increase of IPv6 traffic/prefixes shows a steep increase and
that the application mix of IPv6 traffic starts to imitate the one of
IPv4-dominated Internet.



\end{abstract}

\IEEEpeerreviewmaketitle

\section{Introduction}
\label{sec:introduction}

Although designing a new Internet protocol was not the most pressing matter
within the Internet community in the beginning of the 1990s, the unforeseen
growing speed of the Internet usage had started to cause worries about the
exhaustion of the Internet address space. Despite these concerns, the exhaustion
of the current Internet address space (IPv4) became inevitable due to the
consequence of overstaying in the decision process of the movement to the new
Internet address space (IPv6~\cite{RFC-2460}). Even though the optimistic
prediction of the Internet growth was one of the major mistakes made in the
beginning of its evolution, Internet pioneers are not to be blamed because it
was nearly impossible to expect such a massive success of the Internet at that
point in time.

Internet Protocol Version 6 (IPv6) has been developed by the Internet
Engineering Task Force (IETF) in 1994 with a view to succeeding the current
version of Internet Protocol (IPv4). Besides the expansion of the address space,
developers of IPv6 took several demanding features such as the security (based
on compliance with IPSec), Quality of Service (QoS, via the
prioritization scheme and the non-fragmentation principle), and the
extensibility (using the chain header) into the design consideration, while
maintaining the simplicity of its predecessor (IPv4). Even with such promising
functionalities network operators and software developers were not motivated
enough to adopt the new version of the Internet protocol because the current
version of the Internet protocol is irreproachable and it was still too early
to feel the scarcity of the address space in their bones. Furthermore, it was
commonly acknowledged that inequalities of the benefit of and the demand for
the new Internet protocol among network operators made it difficult to adopt
IPv6 all together.

However, the situation has changed since the last block of the IPv4 addresses
has been assigned to the RIRs in mid-2011. Soon after the IPv4 address
depletion of the Internet Assigned Numbers Authority (IANA), Asia Pacific
Network Information Centre (APNIC) reported that they reached the final stage
of the IPv4 exhaustion~\cite{URL-APNIC-IPV4}. After all, major network
operators nodded at an implicit agreement that there is no more time to
calculate gains and losses. To this end, the Internet Society~\cite{URL-ISOC}
organized a 24-hour global IPv6 test flight (\wid~\cite{URL-V6DAY}) on 8 June,
2011 and more than a thousand globally influential service providers have
participated in the event. As no severe problems have been reported from
participants, the community has decided to move a step forward, which is the
\wil~\cite{URL-V6LAUNCH} event held on 6 June, 2012. The \wil event was
expected to be an important turning point in the Internet history, because
participants agreed to keep their IPv6 connectivity permanently enabled after the
event.

In this paper, we study the changes to the IPv6 traffic from the viewpoint
of a large European Internet Exchange Point (IXP). We analyze 14 months worth of
traffic traces which include the two world IPv6 events.

The remainder of this paper is structured as follows. We first illustrate the
vantage point of our measurement and traffic data sets in~\xref{sec:datasets}.
The methodology used for conducting our measurement is described in
\xref{sec:methodology}. Then, we show the characteristics of the current IPv6
traffic observed in our measurement in~\xref{sec:results}. After that, we
select publicly available reports about the current status of the adoption of
IPv6 and summarize them in \xref{sec:reports}. We overview the related work
in~\xref{sec:relatedwork}. Finally, we conclude the paper
in~\xref{sec:conclusion}.




\section{Measurement Environment and Traces}
\label{sec:datasets}

In this section, we give a brief overview of our data sets and describe the
vantage point that our data is collected from.

\subsection{Internet eXchange Point (IXP)}

\begin{figure}[htp]
\begin{center}
	\includegraphics[width=0.8\linewidth]{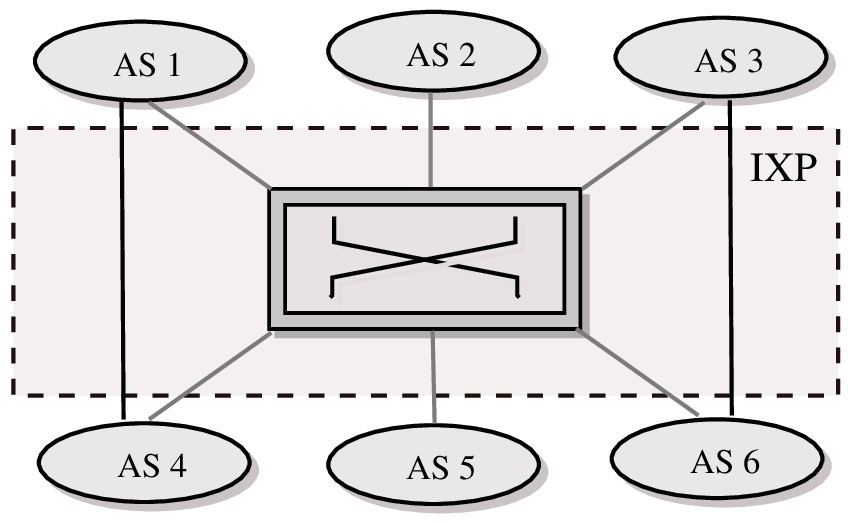}
	\caption{A simplified IXP topology}
	\label{fig:ixp}
\end{center}
\end{figure}

An Internet eXchange Point (IXP) is a physical infrastructure for
interconnecting Internet Service Providers (ISPs) and ASes in order to reduce the
end-to-end latency and the cost of transit traffic. A simplified topology of an IXP
is illustrated in \fref{fig:ixp}. Through such infrastructures, its member ASes can
establish peering (or transit) relationships with other member ASes. The volume
of daily network traffic at an IXP depends, amongst other things, on the number of
networks connected to the IXP. Our IXP has more than 400 members which makes it one
of the largest IXPs in Europe. For a detailed analysis of the IXP ecosystem and the
peering behavior of the IXP participants we refer to~\cite{AOA-SIGCOMM12}. 



\subsection{Data Sets}

\begin{table}
\caption{Data sets.}
\begin{center}
\tabcolsep1mm
\begin{scriptsize}
\begin{tabular}{|l|l|l|}
\hline
Name & Period & Global IPv6 Event \\
\hline
\ixpa	& Jun. 5, 2011 -- Jun. 15, 2011 & \wid \\
\ixpb	& Dec. 25, 2011 -- Jan. 7, 2012  & --- \\ 
\ixpc	& Jun. 1, 2012 -- Jun. 11, 2012 & \wil \\
\ixpd	& Aug. 12, 2012 -- Aug. 30, 2012 & --- \\
\hline
\end{tabular}
\end{scriptsize}
\end{center}
\label{tab:traces}
\end{table}

We base our measurement on four sets of traffic traces (see~\tref{tab:traces})
collected within the above-mentioned IXP during a time span of 14 months
including \wid and \wil (54 days of traffic in total). Note that one of our
data sets (\ixpa) overlaps the one used in Sarrar~\etal~\cite{IIT-PAM12}. Due
to the immense amount of traffic volume, it is considered to be virtually
impossible to capture and store all packets. To this end, we collect our data
using sFlow,~\ie{a network traffic sampling technology
designed to provide a scalable monitoring solution with low costs}, with the
sampling rate of 1:16k. By the reasons of the space constraint and the privacy
concern, we only record the first 128 bytes of the sampled packets. However,
it is sufficient to include all relevant information that we rely on,~\eg{IP headers, tunneling
headers, and transport layer protocol headers}. A sampled packet is pipelined to the anonymization process before
being stored to disk for the purpose of muddling all IP addresses of the
packet, while the consistency of an IP address and the size of the network
prefix are preserved.

\section{Methodology}
\label{sec:methodology}

In order to conduct our measurement, we develop an IPv6 traffic analysis tool
to which text-formatted information extracted from sFlow traces is fed. Our
analysis tool is designed to identify IPv6 packets and to extract relevant
information from the identified IPv6 packets.


The tool identifies the transition technology in the current
measurement point (IXP) by observing the packet's IP version field, protocol
number, and port numbers. For this, we use the protocol number 41 as a clear
indicator of 6in4~\cite{RFC-4213} technology and the protocol number 17
combined with the UDP port number 3544 as a clear evidence of
teredo~\cite{RFC-4380} technology. In the same manner, ayiya~\cite{AYI-IETF}
technology can be identified by using the protocol number 17 and the UDP port
number 5072. However, we do not consider the ayiya transition technology in our
study since we observe only the negligible fraction ($<$0.1\perc) of IPv6
traffic over ayiya. Indeed, teredo and 6in4 are the only transition
technologies carrying a considerable amount of IPv6 traffic.




\section{Evaluation}
\label{sec:results}


In this section, we evaluate the change in the level of IPv6
deployment with emphasis on the impact of the two past global IPv6 events
(\wid and \wil).  
In general, we observe that IPv6 accounts for 0.5\perc of the total Internet
traffic in the peak period (\ixpd) of our measurement results.



\subsection{Quantitative Analysis of IPv6 Traffic}

\begin{figure}[t]
	\includegraphics[width=\linewidth]{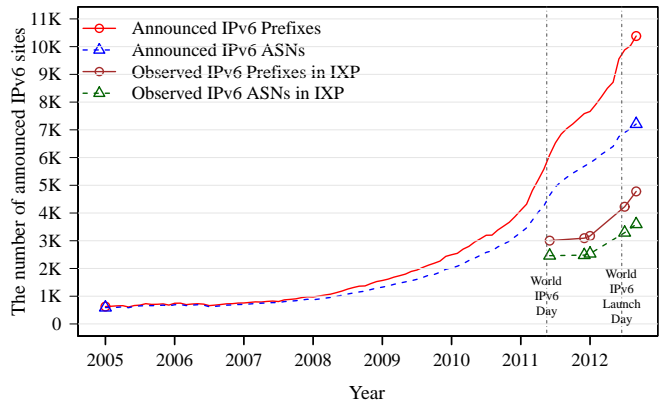}
	\caption{The growth of the IPv6 routing table since 2005 (shorter lines are the observation from our data sets during 14 months of measurement period). Lines from RouteViews data are plotted in monthly intervals and lines from IXP data are plotted five times within the period (each point represents the number of prefixes and ASes observed within the corresponding traffic data set).}
	\label{fig:ipv6rib}
\end{figure}

\begin{figure*}[ht]
	\centering
	\subfigure[IPv6 traffic volume]{
		\includegraphics[width=0.45\linewidth]{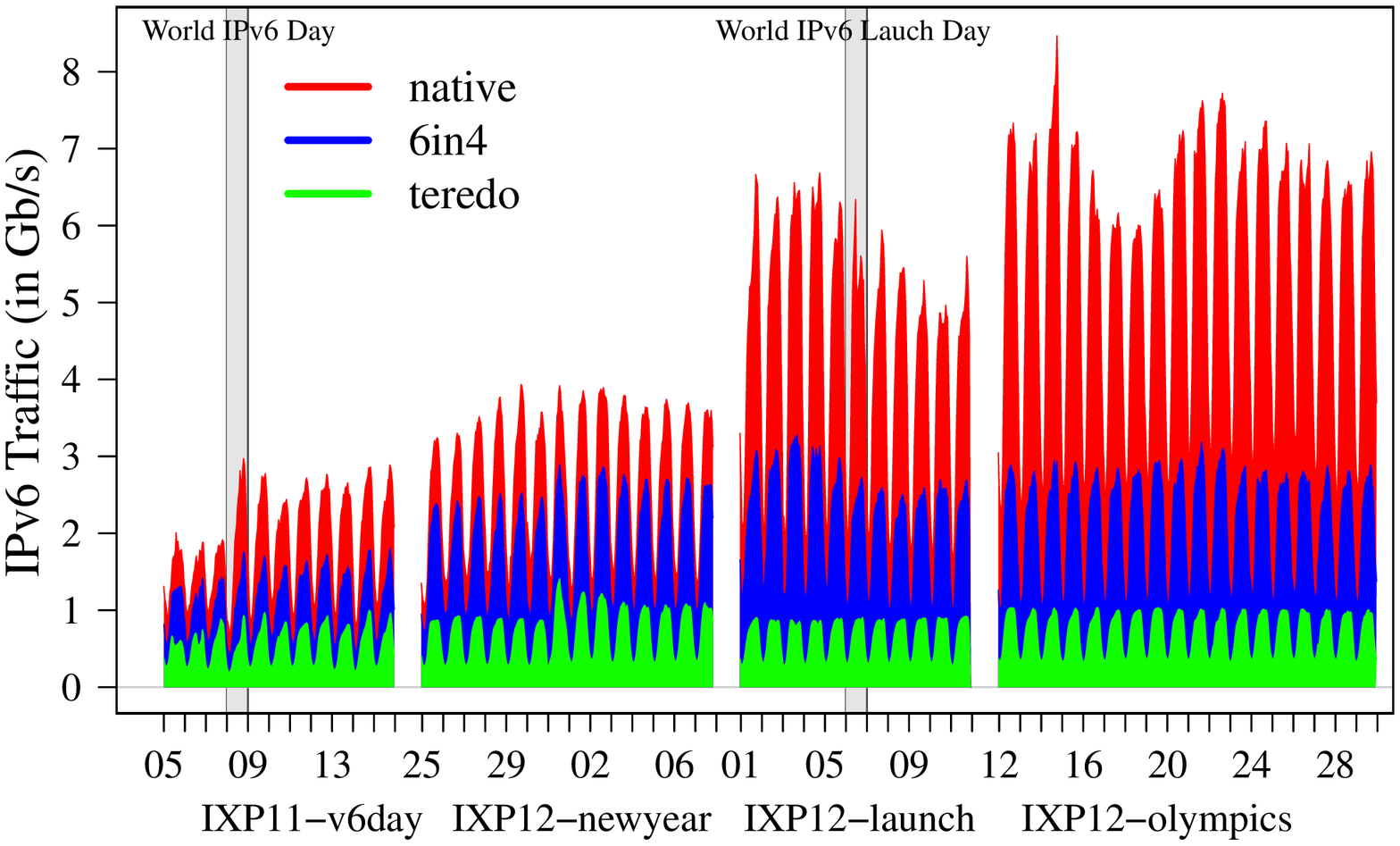}}
	\hspace{25pt}
	\subfigure[IPv6 packet counts]{
		\includegraphics[width=0.45\linewidth]{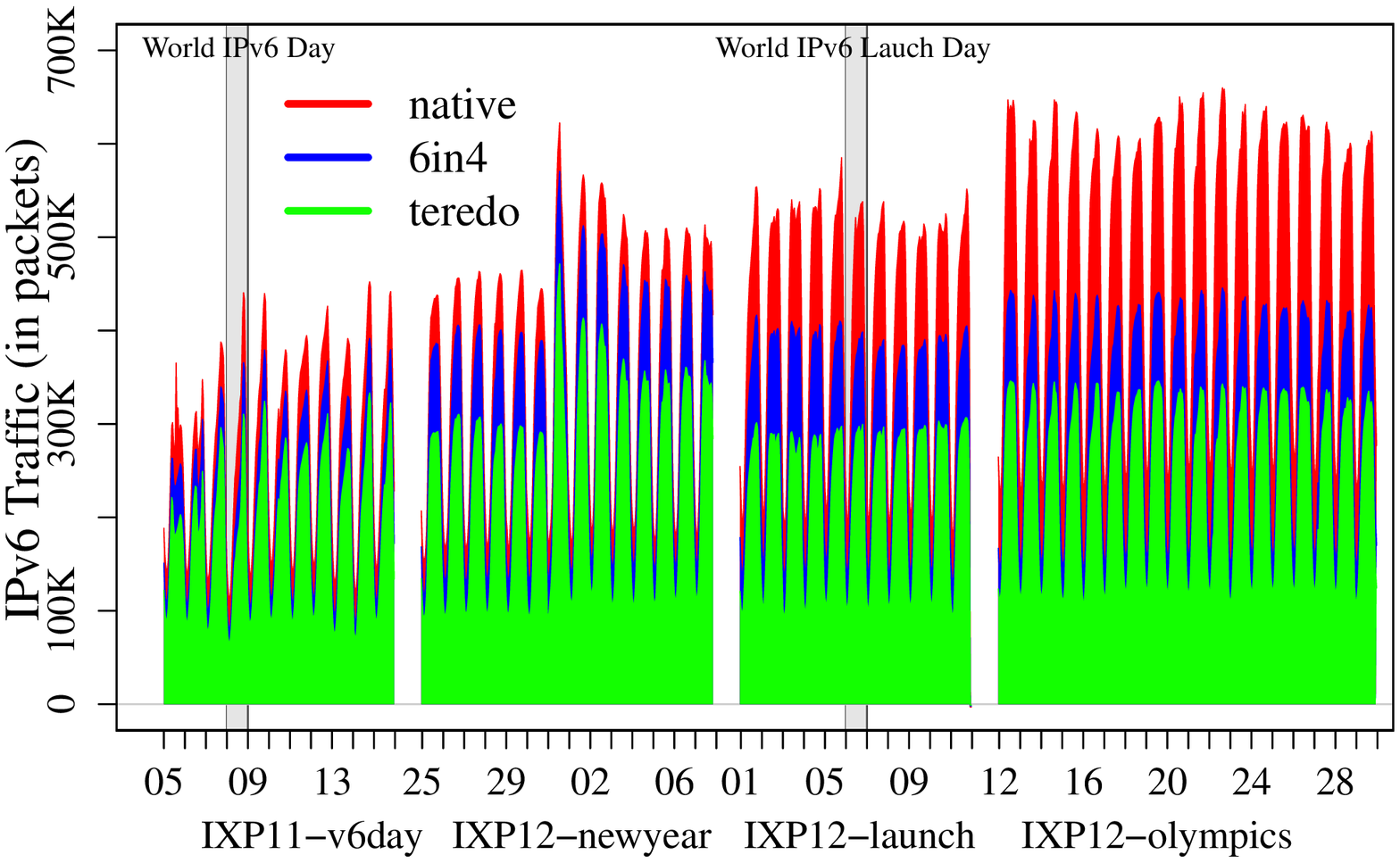}}
	\caption{IPv6 traffic changes.}
	\label{fig:ipv6change}
\end{figure*}

In order to see how the level of IPv6 adoption has been evolving, we download
monthly snapshots of the IPv6 routing table from the RouteViews Project
\cite{URL-ROUTEVIEWS} and illustrate them in~\fref{fig:ipv6rib} together with
the observed IPv6 prefixes within our data sets. The figure shows the sharp rise
in the number of announced IPv6 network prefixes since the \wid. More precisely,
the number of IPv6 network prefixes announced between the \wid period and the
\wil period (one year) is almost equivalent to the total number of prefixes
evolved during the last two decades. Even though we observe only half of the
globally announced IPv6 prefixes at our vantage point, the increasing rate is
still significant.

When seeing the change of the IPv6 adoption in the perspective of the traffic
volume, the increase of IPv6 traffic is even more drastic than the one shown in
the prefix data. As we see in~\fref{fig:ipv6change}, IPv6 traffic is
increased by more than 50\perc on the \wid. Moreover, the volume of IPv6
traffic is again doubled (approximately from 3Gb/s to 6Gb/s) within the \wil
period. The growth of IPv6 traffic is still observed in our latest data set
(\ixpd).

The most significant change that we can discover from the figure is that the
increase of IPv6 traffic mainly results from native IPv6 traffic. On the \wid,
we see more than twice the volume of native IPv6 traffic and it does not decrease after
the event (even though it was supposed to be a 24-hour test flight). Besides,
another multi-fold increase of native IPv6 traffic is experienced in the \wil
period.

With regard to the sudden peak of teredo packets observed on 1 January, 2012
(see \fref{fig:ipv6change} (b)), we do not have a clear answer for this
phenomenon. However, we are of the strong belief that the cause of this peak is
rather due to academic/industrial experiments than due to the participation in
the \wil event. This assumption is based on three different observations. First, the
number of teredo packets falls back to the normal level in the \wil period.
Second, more than 40\perc of the total teredo packets within the period are
generated from a few IP addresses within the same IP prefix. Third, most of the
teredo packets are bubble packets\footnote{A teredo bubble is a signaling
packet typically used for creating and maintaining a NAT mapping} in which
there is no actual payload present.

\begin{figure}[ht]
	\includegraphics[width=\linewidth]{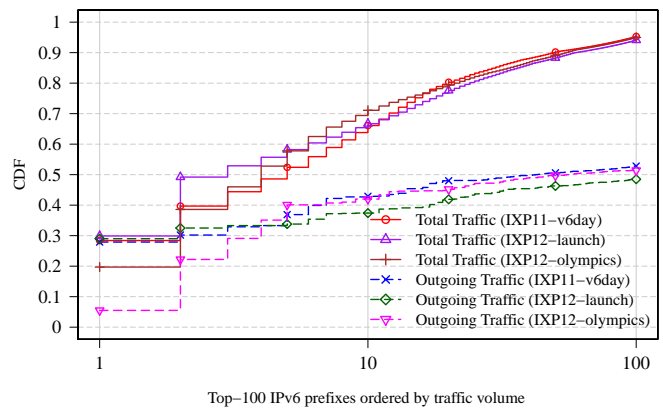}
	\caption{CDF of IPv6 traffic contributed by the top-100 prefixes. The prefixes on the x-axis are sorted by the IPv6 traffic contribution from left to right in descending order. A logarithmic scale is used on the x-axis to clearly identify points of the top prefixes.}
	\label{fig:pfxcdf}
\end{figure}

Next, we evaluate how IPv6 traffic is distributed across prefixes.
\fref{fig:pfxcdf} shows a Cumulative Distribution Function (CDF) of IPv6
traffic contributed by the top-100 IPv6 prefixes. Solid lines in the figure
indicate the traffic fraction of the prefix out of the total observed IPv6
traffic, while dashed lines describe only the fraction of outgoing IPv6
traffic. The figure shows that about 1\perc (50 prefixes in \ixpa and 60
prefixes in \ixpc and \ixpd) of the observed prefixes contribute more than
90\perc of the total IPv6 traffic. Moreover, almost 30\perc (\ixpa and \ixpc)
and 20\perc (\ixpd) of the total traffic is contributed by a single prefix.
Given the fact that almost all traffic of the prefix is outgoing traffic in
\ixpa and \ixpc, we conclude (and also verify) that the top IPv6 contributor
in these periods is a content provider. However, by observing the ratio
between incoming traffic and outgoing traffic of the top prefix in \ixpd, it
appears likely that the first position in the traffic ranking is now taken
by a large transit network. Indeed, by further dataset inspections we confirm
the following assumption: The
identified transit network is one of the largest IPv6 backbone network in the
world.

Yet, readers must note that this evaluation is based on network prefixes. When
prefixes are aggregated into the AS level, the content provider (the top IPv6
traffic contributor in \ixpa and \ixpc) is still the largest IPv6 traffic
contributor in terms of the traffic volume.


\subsection{Application Breakdown}

\begin{figure}[ht]
	\includegraphics[width=\linewidth]{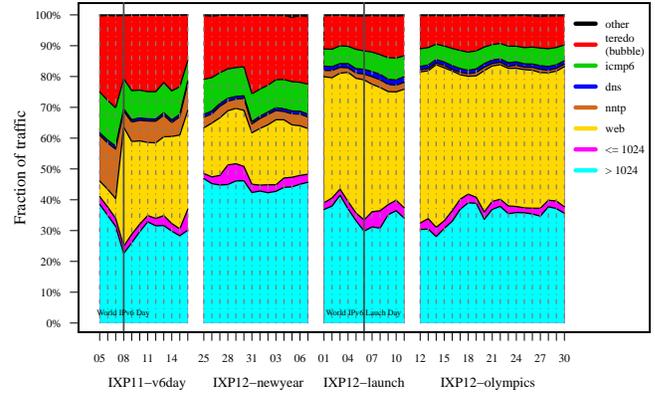}
	\caption{Application breakdown. Traffic volume includes the size of encapsulation headers.}
	\label{fig:trafficbreakdown}
\end{figure}

In this section, we provide insights into the application mix in IPv6
traffic. \fref{fig:trafficbreakdown} shows the breakdown of the traffic into applications.
Note that the header size of transition technologies is included in the
traffic volume. Hence, we see a significant fraction of teredo traffic in
the figure even though teredo bubbles\footnote{A teredo bubble is a signaling
packet typically used for creating and maintaining a NAT mapping} do not contain
any actual payload.

The first application protocol that we need to pay attention to is the Network
News Transport Protocol (NNTP). Even though the significance of NNTP traffic
within the IPv6 network decreases drastically as the influence of web traffic
increases, it is important to understand its characteristics since NNTP has
been the dominant content carrier in the IPv6 world before HTTP became the
major protocol and it is still responsible for a considerable fraction of the
IPv6 traffic.

NNTP has been considered an obsolete protocol for a
while, however recent measurement studies~\cite{TUU-GI10} have found
that NNTP revives (or survives) from its oblivion. They report that NNTP
accounts for up to 5\perc of the total residential traffic in today's Internet.

More surprisingly, Sarrar~\etal~\cite{IIT-PAM12} report that almost 40\perc of
the total IPv6 traffic is contributed by NNTP before the \wid. Our result
shows a significantly different fraction (about 20\perc) of NNTP traffic in the
same period, but this is due to the fact that we consider the header bytes of the
transition technologies as part of the application's traffic volume. When considering
only the payload, our evaluation matches the one reported in~\cite{IIT-PAM12}.


The next application protocol that we study is HTTP. As the figure shows,
the fraction of web traffic increases remarkably on the \wid and it reaches
nearly 50\perc of the total IPv6 traffic in the \wil period. The change of the
traffic shape has a particular meaning since the breakdown of IPv6 traffic
into applications becomes similar to the one of today's IPv4 traffic.
Furthermore, the major share of IPv6 traffic is real content rather than
signaling traffic. This leads us to the conclusion that service providers start
to break away from the fixed idea that IPv6 is a faraway story and become more
active in providing Internet access through IPv6 to their home and enterprise customers. 

\subsection{Native IPv6 Traffic Among ASes}
\begin{figure}[htp]
	\includegraphics[width=\linewidth]{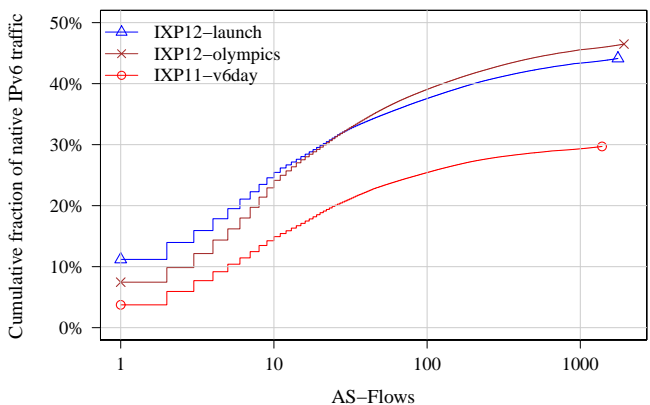}
	\caption{Cumulative fraction of native IPv6 traffic from an AS to another AS (AS-flow). Only AS-flows contributing more than 0.001\perc of total native IPv6 traffic are considered. Links are sorted from left to right by the amount of traffic in descending order.}
	\label{fig:trafficperlink}
\end{figure}

We now investigate how native IPv6 traffic is exchanged among ASes. For this
study, we create a traffic matrix with sorted ASes as columns and rows
(more than 3,500 ASes). \fref{fig:trafficperlink} depicts the cumulative
fraction of native IPv6 traffic of AS-flows~\footnote{We define
an AS-flow as an asymmetric pair of ASes}. In the figure, we only
consider AS-flows which contribute at least 0.001\perc of the total native IPv6
traffic,~\ie{1,386 AS-flows in \ixpa, 1,761 AS-flows in \ixpc, and 1,928
AS-flows in \ixpd}. The figure describes that the traffic fraction of AS-flows
increases remarkably in the \wil period. Interestingly, the amount of IPv6
traffic is largely concentrated in one AS-flow (\ie{the AS-flow numbered as 1
on the x-axis)}. However, the traffic contribution of this top AS-flow
decreases from 11.19\perc (\ixpc) to 7.46\perc (\ixpd), while the traffic share
of those in the bottom 98\perc of \ixpd outperforms that of \ixpc. From this
result, we can infer that the IPv6 traffic relationship among ASes slowly
changes from a 1-n shape to a n-n shape. This phenomenon is likely related to
the rapid growth of the transit network explained in reference to~\fref{fig:pfxcdf}.

\begin{figure*}[htp]
	\centering
	\subfigure[\ixpa]{
		\includegraphics[width=0.24\linewidth]{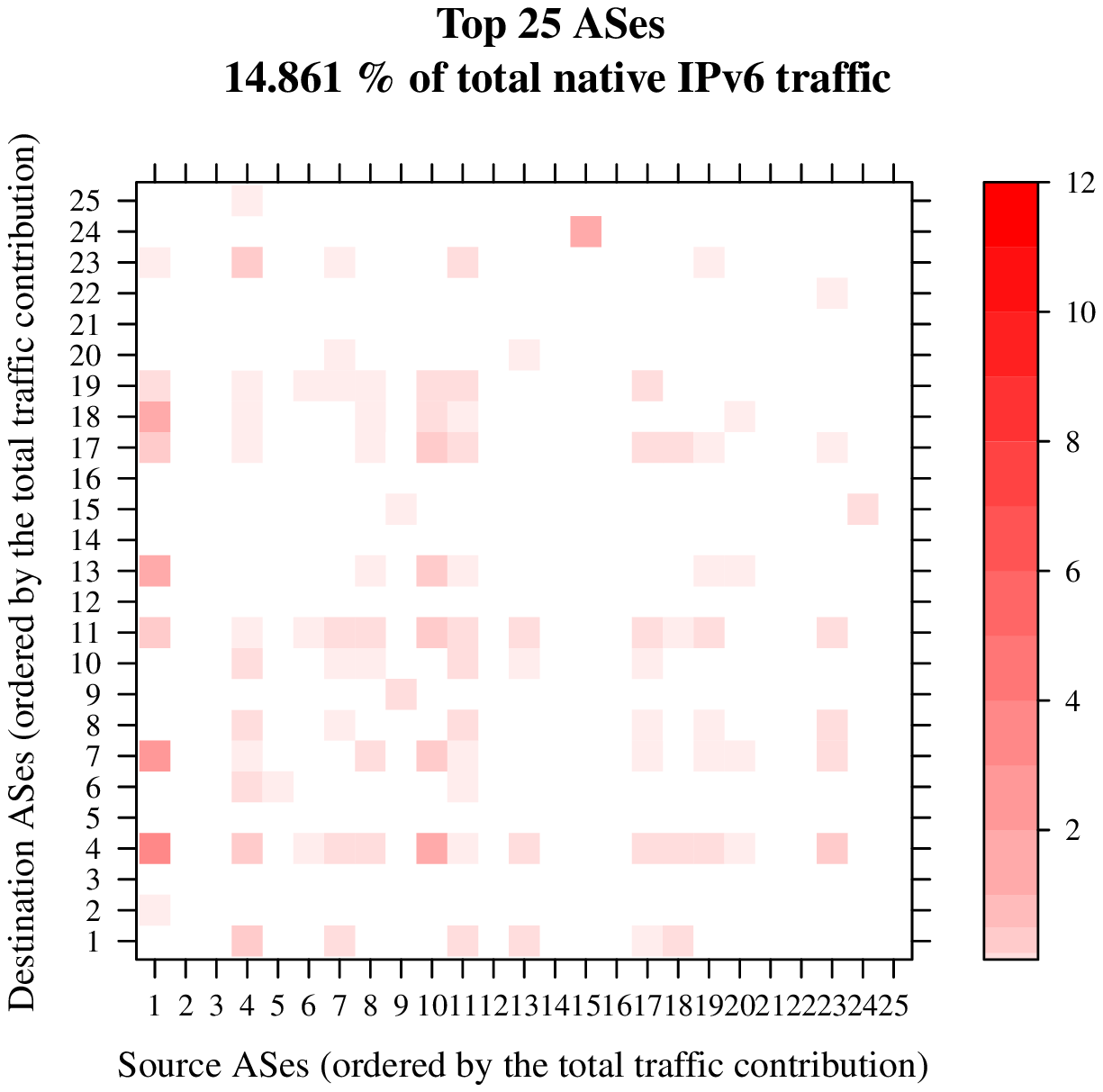}}
	\hspace{30pt}
	\subfigure[\ixpc]{
		\includegraphics[width=0.24\linewidth]{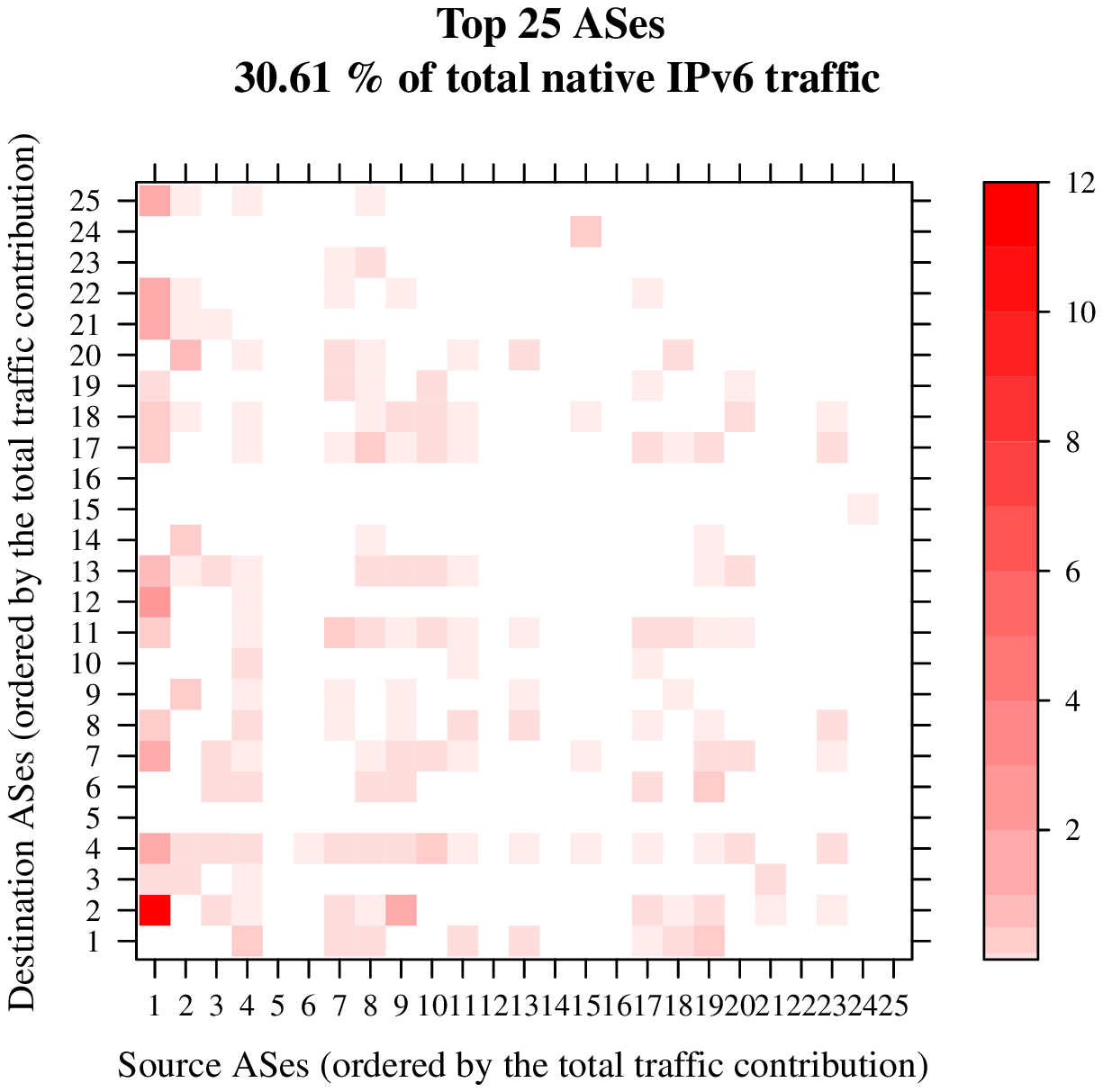}}
	\hspace{30pt}
	\subfigure[\ixpd]{
	    \includegraphics[width=0.24\linewidth]{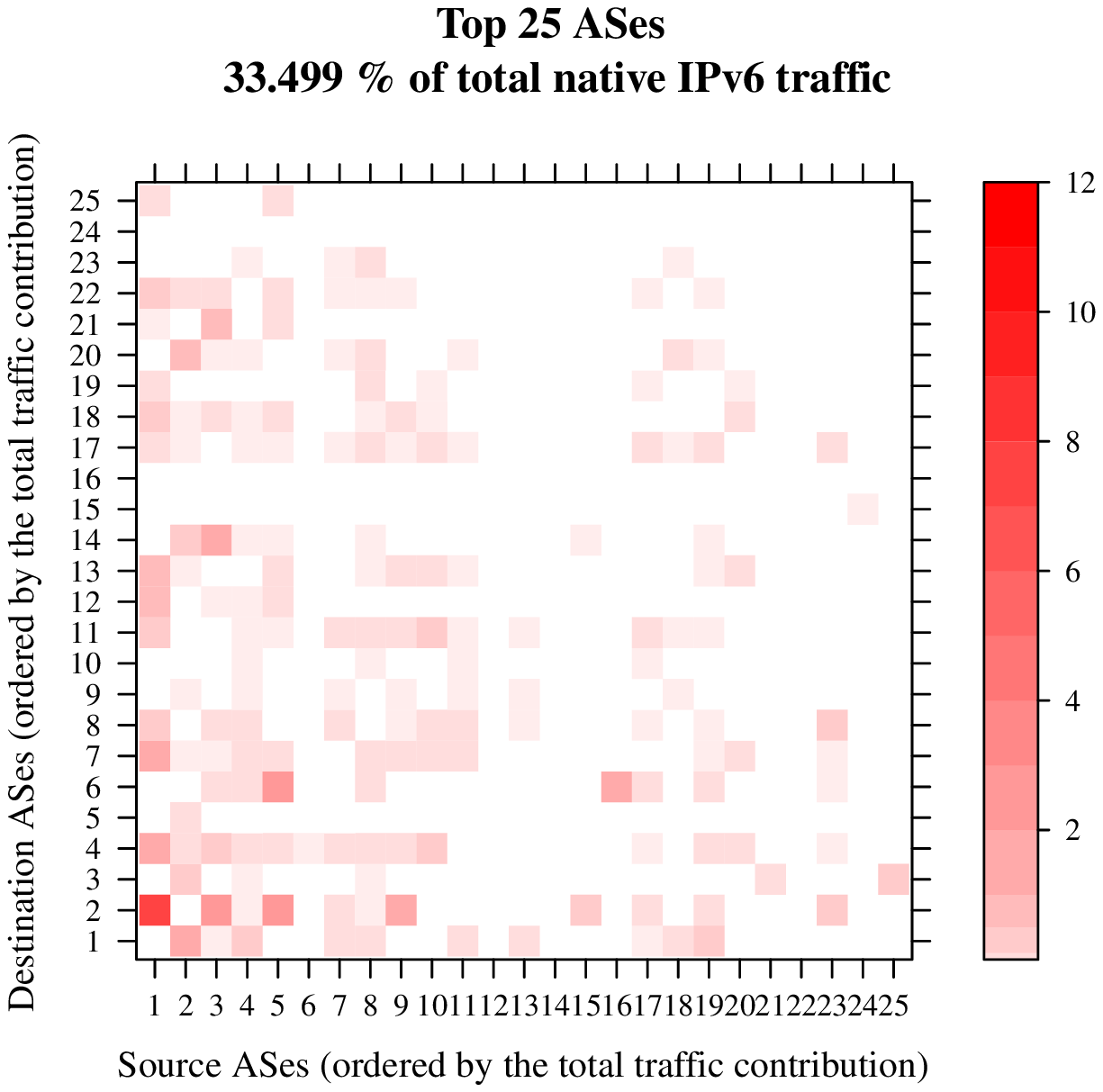}}
	\caption{Traffic among top-25 ASes. ASes are sorted from left to right on the x-axis (from bottom to top on the y-axis) by the volume of contributing native IPv6 traffic in descending order.} 
	\label{fig:trafficmatrix}
\end{figure*}

Before providing more information of this specific AS-flow, we narrow the scope
of our investigation. For doing so, we illustrate the matrix of the native IPv6
traffic fraction among the top-25 ASes in \fref{fig:trafficmatrix}. The cell in
the figures are filled with different levels of color depth according to their
share in the total native IPv6 traffic. Sums of IPv6 traffic transmitted among
these top-25 ASes are 14.86\perc, 30.61\perc, and 33.50\perc in \ixpa, in
\ixpc, and in \ixpd, respectively. Readers might notice by the intuition that
the cell filled with the deepest color in \fref{fig:trafficmatrix} (b) and
\fref{fig:trafficmatrix} (c),~\ie{ones representing the fraction of traffic
originating from the AS numbered as 1 (x-axis) and destining for the AS numbered
as 2 (y-axis)}, are corresponding to the largest AS-flow discussed in regard
to~\fref{fig:trafficperlink}. Furthermore, we can determine that the AS numbered
as 1 is the major source of IPv6 contents. Adding up all native IPv6 traffic
generated from this specific AS, we find that 15.21\perc (\ixpa), 29.93\perc
(\ixpc), and 17.88\perc (\ixpd) originate from that single AS.

From a closer inspection of that AS, we identify that this AS belongs to one
of the largest content providers in the world. This might not be an eye-opener
for readers because the consumer behavior of Internet contents in the IPv6
world cannot be very different from the one within the IPv4 world.
However, we believe that by studying these trends in IPv6 traffic and their core content
providers we can help network operators to come up with more effective
strategies for implementing IPv6 in their networks.

Readers may still wonder about the identity of the AS in which the most IPv6
traffic from the top content provider is consumed (the AS numbered as 2 in
\fref{fig:trafficmatrix}). Regarding the largest AS-flow which is shown in
\fref{fig:trafficperlink} and is also illustrated as the cell filled with the
deepest color in~\fref{fig:trafficmatrix}, we reveal that the top IPv6 traffic
consumer in our measurement is one of the largest ISPs located in Romania.
This is a somewhat unexpected discovery for us since Romania has not engaged
any appreciable attentions in the Internet history before. Taking a closer look
at the traffic of this specific ISP in the peak period of our IPv6 traffic
(\ixpd), approx. 3.02\perc of the total native IPv6 traffic is going out from
this AS, while about 14.23\perc of the total traffic is going into the AS.





\section{Public Reports on IPv6 Traffic}
\label{sec:reports}

Various content providers and service providers have been reporting the status
of IPv6 traffic shown in their networks. In this section, we select three
relevant reports and summarize them in relation to our measurement results.

\subsection{Google}
Google has been collecting the statistics about the IPv6 connectivity of their
users and reporting the results since 2008~\cite{URL-GOOGLE}. They claim that
0.34\perc of the total users have accessed their website over IPv6 on the \wid
and the fraction has increased to 0.65\perc on the \wil. Given the fact that
Google is the top ranked website in terms of the number of visitors and the
traffic volume, these fractions indicate a considerable number of users.
Our measurement witnesses that Google is one of the biggest sources of IPv6
traffic.

Another observation they have made is that since March 2010 the
fraction of IPv6 users of the native IPv6 technology has overtaken the one of
encapsulation technologies, \ie{teredo and 6in4}. According to their report,
the tendency of the decrease in users of transition technologies becomes more
drastic over time. As a result, the share of users behind transition
technologies decreases from 11.76\perc (\wid) to 1.54\perc (\wil) of the total
IPv6 users. Although our observation is based on the traffic volume, we confirm
the tendency of the native IPv6 domination. More precisely, we see that the
fraction of IPv6 traffic transferred with transition technologies decreases
from 55.49\perc (\wid) to 46.74\perc (\wil). From our measurement viewpoint,
the amount of native IPv6 traffic has begun to outstrip the amount of IPv6
traffic via transition technologies since the \wid.



\subsection{Hurricane Electric}
Hurricane Electric is a global Internet backbone and one of the largest
networks in terms of the number of customers as well as the
largest IPv6 network in terms of the number of connected networks. According to
the report of Hurricane Electric~\cite{URL-HE}, in the middle of 2012,
85.4\perc of the global Top Level Domains 
(TLDs) have IPv6 name servers. In an investigation of the top 1,000 Usenet
servers, they find out that 16.22\perc of them have IPv6 addresses. Given the
fact that the Network News Transport Protocol (NNTP, the protocol used for
Usenet) accounts for up to 5\perc of the residential network
traffic~\cite{TUU-GI10}, the reported number of IPv6-enabled Usenet servers may
produce a substantial fraction of IPv6 traffic. Indeed,
Sarrar~\etal~\cite{IIT-PAM12} report that almost 40\perc of the total IPv6 is
NNTP traffic before \wid. We also observe a considerable amount of NNTP traffic
from our measurement.

\subsection{APNIC}
The Asia Pacific Network Information Centre (APNIC) is the first regional
Internet registry whose IPv4 address pool has been exhausted. APNIC measures the
deployment level of IPv6 based on the IPv6 preference for IPv4 addresses using
the BGP data. APNIC's report describes that Europe has the highest IPv6
preference (0.51 in July, 2012) for IPv4 addresses among all continents. With
the same metric, they report that Romania is the highest ranked country in the
world. Similarly, the snapshot of Google's IPv6 statistics in the same period
shows that 8.38\perc of their users from Romania access the website using IPv6
which is the highest number among all countries. We confirm that more than
17\perc of the total IPv6 traffic in the peak period of our measurement is
contributed by a Romanian ISP.

\section{Related Work}
\label{sec:relatedwork}

Most of the studies made before the announcement of the name space exhaustion 
focused mainly on performance analyses of various transition techniques
\cite{EIT-CIT03}, IPv4 vs IPv6 comparison~\cite{AIT-CONEXT11}, and the
deployment level of IPv6 based on the address reachability~\cite{OII-PAM08,
QTE-PAM09,EIA-PAM10}. This tendency, however, became less pronounced as time
passes. Instead, researchers start to take a close interest in the
characteristics of IPv6 traffic based on real-world traces~\cite{WGO-NOMS12,
IIT-PAM12}. In this section, we choose a small number of publications which
also studied IPv6 and briefly discribe their work.

Raicu~\etal~\cite{EIT-CIT03} evaluated and compared the performance of the two
different IPv4-to-IPv6 transition mechanisms,~\ie host-to-host encapsulation
mechanism (6-over-4) and router-to-router tunneling mechanism
(6in4), in terms of TCP latency, throughput, and CPU utilization. 

Malone~\etal~\cite{OII-PAM08} quantified the number of IPv6 addresses accessable
in the Internet based on traffic data collected within specific websites and DNS
servers. They used prefixes of identified IPv6 addresses in order to classify
IPv4-to-IPv6 transition techniques. During their measurement (mid. 2006), the
authors observed the remarkable increase of \teredo prefixes, but still the least
used compared to other transition mechanisms, within their data sets. Karpilovsky
~\etal~\cite{QTE-PAM09} also measured the number of publicly announced IPv6
prefixes by analyzing snapshots of BGP data obtained from RouteViews Project.
They reported that RIPE,~\ie Regional Internet Registry (RIR) for European
countries, is the increasingly dominant registrar for IPv6 address allocation.
Colitti~\etal~\cite{EIA-PAM10} analyzed passively collected data from Google's
IPv6-enabled web page and evaluated the degree of the IPv6 adoption. Their
evaluation shows that the adoption of IPv6 is still low but steadily growing and
that the vast majority of IPv6 traffic is contributed by only small number of
networks.


Gao~\etal~\cite{WGO-NOMS12} proposed the method to classify the P2P traffic
from flow-baed IPv6 traffic and evaluated the stage of the IPv6 deployment in
China. They found that P2P and streaming applications are accounting for the
major fraction of IPv6 traffic in China.

Nikkhah~\etal~\cite{AIT-CONEXT11} assessed the performance of IPv6 and compare
it to the performance of IPv4. They claimed that the inefficient pathfinding is
the major cause of the poor performance that IPv6 shows.

Sarrar~\etal~\cite{IIT-PAM12} observed changes of IPv6 traffic on \wid by
analyzing traffic traces collected from the same monitoring point that our study
bases on. They reported the significant increase of IPv6 traffic (more than twice)
on the event day and found that IPv6 traffic did not decrease even after the end
of the event. 


\section{Conclusions}
\label{sec:conclusion}

In this paper, we evaluated the current status of IPv6 traffic by analyzing the
traffic trace collected from a large European IXP during 14 months of the
time span including the two global IPv6 events (\wid and \wil).




Even though we observed that IPv6 traffic accounts for only small
fraction of the total Internet traffic (0.5\perc in the peak period within our
measurement), our study on IPv6 traffic showed that the deployment of IPv6
technology is finally on the right track and we are slowly overcoming the shyness
facing the new Internet technology. Our claim is based on three factors we have
shown in the paper.~\first{the sharp rise in the number of IPv6 prefixes and the
IPv6 traffic has been observed since the \wid and the increase rate does not slow
down after the the \wil event.}~\second{the application mix of the IPv6 traffic
began to form the current traffic status of the IPv4-dominated Internet,
\eg{about 50\perc of HTTP traffic}.}~\third{the fraction of native IPv6 traffic
overtook that of traffic transferred within IPv6-over-IPv4 technologies.} We are
of the belief that continuous efforts on the IPv6 deployment such as \wid and
\wil will keep this increasing tendency.

\bibliographystyle{IEEEtran}
\bibliography{IEEEtran}

\begin{thebibliography}{10}
\providecommand{\url}[1]{#1}
\csname url@samestyle\endcsname
\providecommand{\newblock}{\relax}
\providecommand{\bibinfo}[2]{#2}
\providecommand{\BIBentrySTDinterwordspacing}{\spaceskip=0pt\relax}
\providecommand{\BIBentryALTinterwordstretchfactor}{4}
\providecommand{\BIBentryALTinterwordspacing}{\spaceskip=\fontdimen2\font plus
\BIBentryALTinterwordstretchfactor\fontdimen3\font minus
  \fontdimen4\font\relax}
\providecommand{\BIBforeignlanguage}[2]{{%
\expandafter\ifx\csname l@#1\endcsname\relax
\typeout{** WARNING: IEEEtran.bst: No hyphenation pattern has been}%
\typeout{** loaded for the language `#1'. Using the pattern for}%
\typeout{** the default language instead.}%
\else
\language=\csname l@#1\endcsname
\fi
#2}}
\providecommand{\BIBdecl}{\relax}
\BIBdecl

\bibitem{URL-V6DAY}
\BIBentryALTinterwordspacing
Archive: 2011 world ipv6 day. [Online]. Available:
  \url{http://www.internetsociety.org/ipv6/archive-2011-world-ipv6-day}
\BIBentrySTDinterwordspacing

\bibitem{URL-V6LAUNCH}
\BIBentryALTinterwordspacing
World ipv6 launch. [Online]. Available: \url{http://www.worldipv6launch.org}
\BIBentrySTDinterwordspacing

\bibitem{RFC-2460}
\BIBentryALTinterwordspacing
S.~Deering and R.~Hinden, ``{Internet Protocol, Version 6 (IPv6)
  Specification},'' RFC 2460 (Draft Standard), Internet Engineering Task Force,
  Dec. 1998, updated by RFCs 5095, 5722, 5871. [Online]. Available:
  \url{http://www.ietf.org/rfc/rfc2460.txt}
\BIBentrySTDinterwordspacing

\bibitem{URL-APNIC-IPV4}
\BIBentryALTinterwordspacing
Apnic's ipv4 pool usage. [Online]. Available:
  \url{http://www.apnic.net/community/ipv4-exhaustion/graphical-information}
\BIBentrySTDinterwordspacing

\bibitem{URL-ISOC}
\BIBentryALTinterwordspacing
The internet society. [Online]. Available: \url{http://www.internetsociety.org}
\BIBentrySTDinterwordspacing

\bibitem{AOA-SIGCOMM12}
\BIBentryALTinterwordspacing
B.~Ager, N.~Chatzis, A.~Feldmann, N.~Sarrar, S.~Uhlig, and W.~Willinger,
  ``Anatomy of a large european ixp,'' in \emph{Proceedings of the ACM SIGCOMM
  2012}, Aug. 2012. [Online]. Available:
  \url{http://www.net.t-labs.tu-berlin.de/papers/ACFSUW-ALEI-12.pdf}
\BIBentrySTDinterwordspacing

\bibitem{IIT-PAM12}
N.~Sarrar, G.~Maier, B.~Ager, R.~Sommer, and S.~Uhlig, ``Investigating ipv6
  traffic: what happened at the world ipv6 day?'' in \emph{Proceedings of the
  13th international conference on Passive and Active Measurement}, ser.
  PAM'12.\hskip 1em plus 0.5em minus 0.4em\relax Berlin, Heidelberg:
  Springer-Verlag, 2012, pp. 11--20.

\bibitem{RFC-4213}
\BIBentryALTinterwordspacing
E.~Nordmark and R.~Gilligan, ``{Basic Transition Mechanisms for IPv6 Hosts and
  Routers},'' RFC 4213 (Proposed Standard), Internet Engineering Task Force,
  Oct. 2005. [Online]. Available: \url{http://www.ietf.org/rfc/rfc4213.txt}
\BIBentrySTDinterwordspacing

\bibitem{RFC-4380}
\BIBentryALTinterwordspacing
C.~Huitema, ``{Teredo: Tunneling IPv6 over UDP through Network Address
  Translations (NATs)},'' RFC 4380 (Proposed Standard), Internet Engineering
  Task Force, Feb. 2006, updated by RFCs 5991, 6081. [Online]. Available:
  \url{http://www.ietf.org/rfc/rfc4380.txt}
\BIBentrySTDinterwordspacing

\bibitem{AYI-IETF}
\BIBentryALTinterwordspacing
J.~Massar, ``{AYIYA: Anything In Anything},'' Internet Engineering Task Force,
  Jan. 2005. [Online]. Available:
  \url{http://tools.ietf.org/html/draft-massar-v6ops-ayiya-02}
\BIBentrySTDinterwordspacing

\bibitem{URL-ROUTEVIEWS}
\BIBentryALTinterwordspacing
Routeviews project. [Online]. Available: \url{http://www.routeviews.org}
\BIBentrySTDinterwordspacing

\bibitem{TUU-GI10}
J.~Kim, F.~Schneider, B.~Ager, and A.~Feldmann, ``Today's usenet usage:
  Characterizing nntp traffic,'' in \emph{Proceedings of INFOCOM IEEE
  Conference on Computer Communications Workshops}.\hskip 1em plus 0.5em minus
  0.4em\relax New York, NY, USA: IEEE, March 2010, p. 1–6.

\bibitem{URL-GOOGLE}
\BIBentryALTinterwordspacing
Google ipv6 statistics. [Online]. Available:
  \url{http://www.google.com/ipv6/statistics.html}
\BIBentrySTDinterwordspacing

\bibitem{URL-HE}
\BIBentryALTinterwordspacing
Global ipv6 deployment progress report. [Online]. Available:
  \url{http://bgp.he.net/ipv6-progress-report.cgi}
\BIBentrySTDinterwordspacing

\bibitem{EIT-CIT03}
I.~Raicu and S.~Zeadally, ``Evaluating ipv4 to ipv6 transition mechanisms,'' in
  \emph{Proceedings of the 10th International Conference on
  Telecommunications}, vol.~2, Feb. 2003, pp. 1091 -- 1098 vol.2.

\bibitem{AIT-CONEXT11}
\BIBentryALTinterwordspacing
M.~Nikkhah, R.~Gu{\'e}rin, Y.~Lee, and R.~Woundy, ``Assessing ipv6 through web
  access a measurement study and its findings,'' in \emph{Proceedings of the
  7th COnference on emerging Networking EXperiments and Technologies}, ser.
  CoNEXT'11.\hskip 1em plus 0.5em minus 0.4em\relax New York, NY, USA: ACM,
  2011, pp. 26:1--26:12. [Online]. Available:
  \url{http://doi.acm.org/10.1145/2079296.2079322}
\BIBentrySTDinterwordspacing

\bibitem{OII-PAM08}
\BIBentryALTinterwordspacing
D.~Malone, ``Observations of ipv6 addresses,'' in \emph{Proceedings of the 9th
  international conference on Passive and active network measurement}, ser.
  PAM'08.\hskip 1em plus 0.5em minus 0.4em\relax Berlin, Heidelberg:
  Springer-Verlag, 2008, pp. 21--30. [Online]. Available:
  \url{http://dl.acm.org/citation.cfm?id=1791949.1791953}
\BIBentrySTDinterwordspacing

\bibitem{QTE-PAM09}
\BIBentryALTinterwordspacing
E.~Karpilovsky, A.~Gerber, D.~Pei, J.~Rexford, and A.~Shaikh, ``Quantifying the
  extent of ipv6 deployment,'' in \emph{Proceedings of the 10th international
  conference on Passive and Active Measurement}, ser. PAM'09.\hskip 1em plus
  0.5em minus 0.4em\relax Berlin, Heidelberg: Springer-Verlag, 2009, pp.
  13--22. [Online]. Available:
  \url{http://dx.doi.org/10.1007/978-3-642-00975-4_2}
\BIBentrySTDinterwordspacing

\bibitem{EIA-PAM10}
\BIBentryALTinterwordspacing
L.~Colitti, S.~Gunderson, E.~Kline, and T.~Refice, ``Evaluating ipv6 adoption
  in the internet,'' in \emph{Proceedings of the 11th international conference
  on Passive and Active Measurement}, ser. PAM'10.\hskip 1em plus 0.5em minus
  0.4em\relax Berlin, Heidelberg: Springer-Verlag, 2010, pp. 141--150.
  [Online]. Available: \url{http://dl.acm.org/citation.cfm?id=1889324.1889339}
\BIBentrySTDinterwordspacing

\bibitem{WGO-NOMS12}
L.~Gao, J.~Yang, H.~Zhang, D.~Qin, and B.~Zhang, ``What's going on in chinese
  ipv6 world,'' in \emph{Network Operations and Management Symposium (NOMS),
  2012 IEEE}, Apr. 2012, pp. 534 --537.

\end{thebibliography}

\end{document}